# Synergy, not size: How collaboration architecture shapes scientific disruption

Bili Zheng[1], Jianhua Hou[1]

1 School of Information Management, Sun Yat-sen University, Guangzhou, 51006, China

Abstract: The mechanisms driving different types of scientific innovation through collaboration remain poorly understood. Here we develop a comprehensive framework analyzing over 14 million papers across 19 disciplines from 1960-2020 to unpack how collaborative synergy shapes research disruption. We introduce the synergy factor $R(g)$ to quantify collaboration cost-benefit dynamics, revealing discipline-specific architectures where Physics peaks at medium team sizes while humanities achieve maximal synergy through individual scholarship. Our mediation analysis demonstrates that collaborative synergy—not team size alone—mediates 75-85% of the relationship between team composition and disruption. Key authors play a catalytic role, with papers featuring exceptional researchers showing 561% higher disruption indices. Surprisingly, high-citation authors reduce disruptive potential while those with breakthrough track records enhance it, challenging traditional evaluation metrics. We identify four distinct knowledge production modes—elite-driven, baseline, heterogeneity-driven, and low-cost—each optimized for different research objectives. These findings reveal substantial heterogeneity in optimal collaboration strategies across disciplines and provide evidence-based guidance for research organization, with implications for science policy and the design of research institutions in an increasingly collaborative scientific landscape.

The escalating scale and complexity of scientific collaboration have cemented team-based research as the dominant paradigm for tackling grand challenges and advancing the frontiers of knowledge (Hall et al., 2018). This paradigm shift presents both unprecedented opportunities for discovery and profound questions about how innovation is fostered and knowledge evolves within these collaborative structures (Yao et al., 2025). While considerable research has explored the impact of factors such as team size and interdisciplinarity on scientific output (Lin et al., 2023; Lin et al., 2025; Uzzi et al., 2013; Wu et al., 2019), a more granular understanding of the mechanisms driving different types of innovation—notably disruptive and developmental breakthroughs—remains elusive. Specifically, how the internal architecture of teams, the synergistic dynamics of their collaboration, and the contributions of pivotal individuals interplay to shape these distinct output patterns is not yet fully understood (Liu et al., 2024; Lyu et al., 2021; Yoo et al., 2024).

A mechanistic understanding of these dynamics is crucial for optimizing science policy,

allocating research funding, and cultivating diverse innovation ecosystems (Wu et al., 2024). Seminal work has established an empirical link between team size and the nature of innovation, suggesting that smaller teams are more prone to disrupt existing scientific and technological trajectories with novel ideas, while larger teams tend to develop and consolidate established knowledge domains (Wu et al., 2019). This foundational observation provides a critical entry point but simultaneously raises more fundamental questions about the decision-making of scientists and the emergent outcomes of their interactions: What intrinsic dynamics lead to these divergent outcomes? Is it merely an effect of scale, or are more complex factors at play?

However, conceptualizing teams as undifferentiated collections of individuals or focusing predominantly on their size may obscure the nuanced realities of scientific collaboration (Lin et al., 2025). Scientific creation is not a simple aggregation of effort but rather a complex interplay of knowledge exchange, cost-benefit trade-offs in cooperation, and the heterogeneous contributions of individual members (Kozlowski, 2017; Kozlowski & Ilgen, 2006). In particular, "key authors"—scholars distinguished by exceptional academic standing, a track record of disruptive innovation, unique interdisciplinary expertise, or rich scientific experience—may exert a disproportionate social influence on a team's research trajectory, choice of problems, and ultimately, the disruptive potential of its outputs (Betancourt et al., 2023; Katchanov et al., 2023). These considerations suggest that a comprehensive understanding of scientific innovation necessitates an integrative perspective that moves beyond scale to interrogate team architecture, the game-theoretic underpinnings of cooperation, and the specific effects of core individuals.

Drawing from the evolutionary dynamics of cooperation (Alvarez-Rodriguez et al., 2021; Nowak, 2006; Perc et al., 2017), this study develops and empirically tests a multilevel analytical framework to reveal the structural foundations and synergistic mechanisms of scientific teams' knowledge production modes. We focus on the interplay among three core elements: (1) Adapting the concept of a "Synergy Factor" ($R(g)$) from cooperative game theory to quantify the cost-benefit dynamics of teamwork across varying scales and contexts (Alvarez-Rodriguez et al., 2021); (2) Employing hypergraph theory to characterize the fine-grained higher-order structural features within and between teams (Boccaletti et al., 2023); (3) Innovatively operationalizing and quantifying the "key author effect" to assess the unique leverage of core members on team innovation. Analyzing a large-scale, cross-disciplinary, longitudinal database of academic publications, we aim to identify distinct "knowledge production modes" defined by these interacting factors and to elucidate how these modes account for the observed spectrum of disruptive and developmental scientific outputs. This research seeks not only to offer novel mechanistic explanations for team-based innovation but also to provide actionable insights for fostering effective and diverse scientific collaboration.

**Methods**

**Data sources**

We constructed our base dataset using SciSciNet. To ensure that each paper has at least 5 years of citation data, we collected publication data from 1960 to 2020, covering books, book chapters, conference papers, datasets, journal articles, repositories, and theses. This data was used to build the author-paper relationship. SciSciNet follows the field classification system used by Microsoft Academic Graph, which includes 19 top-level fields and 292 sub-level fields. SciSciNet assigns each paper a top-level field and several sub-level fields. Additionally, for gender classification, SciSciNet uses the genderize.io algorithm to predict the gender of each author and provides both the number of observations and the gender probability for each prediction. Since the subsequent analysis involves disciplines, gender, DI, and other factors, we have retained paper data with valid values for all indicators, totaling 14,330,140 papers.

**Disruption index**

In order to characterize the nature of scientific innovation, the disruption index ($DI$), an indicator based on the citation network, was proposed to measure whether a scientific paper is disruptive or consolidative (Funk & Owen-Smith, 2016; Wu et al., 2019). On this basis, Bornmann & Tekles (2021) proposed $DI_l$, aiming to mitigate the impact of random citations on the results of $DI$. Specifically, for a focal paper (FP), there are three types of citing papers: those citing only the FP (denoted as $i$), those citing only the references of the FP (denoted as $k$), and those citing both the FP and its references (denoted as $j$). The calculation formula for the $DI_l$ of a paper is as follows:

$$DI_l = \frac{N_i - N_j^l}{N_i + N_j^l + N_k} \tag{1}$$

where $N_i$ and $N_k$ represent the number of papers $i$ and $k$, respectively. $N_j^l$ denotes the number of papers $j$ that cite at least $l$ references of the FP. The parameter $l$ serves to provide a threshold for $N_j$, obtaining papers $j$ that are more closely related to the references of FP and thus reducing the impact of random citations. When $l=1$, $DI_l$ is equal to the original $DI$. The value range of $DI_l$ is from -1 (consolidative) to 1 (disruptive). The implications of this indicator and its application scenarios can be found in existing studies (Bornmann & Tekles, 2019, 2021; Park et al., 2023; Yang et al., 2025).

Setting thresholds is a critical step in calculating $DI_l$. First, the scope of the citation window needs to be determined. Bornmann & Tekles (2019) found that a citation window of at least 5 years is more conducive to obtaining stable results. Secondly, the value of $l$ needs to be determined. Bornmann & Tekles (2021) shows that $DI_5$ performs better than the original $DI$. In order to obtain robust results, we set the citation window to 5 years, i.e., calculate the $DI_5$ for each paper based on the citing papers from the publication year *t* of the FP to *t+5* years.

**Synergy factor in collaboration**

We constructed coauthorship hypergraphs for each of the 19 disciplines included in the SciSciNet dataset. In these hypergraphs, nodes represent individual scientists, and hyperlinks represent scientific publications. Each hyperlink connects all coauthors of a given paper and thus encodes a higher-order interaction of order $(g)$, defined as the number of authors on the publication. This hypergraph formalism allows us to capture the structure of group-based collaborations more faithfully than traditional pairwise coauthorship networks. The synergy factor, $R(g)$, is not a direct measurement of the micro-level psychological processes within a team. Instead, it is a macro-level behavioral construct derived from the economic principle of "revealed preferences." The core assumption is that the long-term, aggregate distribution of team sizes $(p_g)$ within a discipline reflects a collective equilibrium reached as scientists implicitly weigh the costs and benefits of different collaborative arrangements. A more frequent team size is thus inferred to offer a more optimal "effective payoff structure." This macro-level approach is a key strength, as it avoids the biases inherent in small-scale surveys and provides a novel, quantifiable, and comparable tool for understanding the "economics of collaboration" across diverse scientific fields.

For each discipline, we extracted the total number of publications (hyperlinks) $L_g$ for each group size $g$. $N$ denotes the total number of unique authors within that discipline. Following the framework of Alvarez-Rodriguez et al. (2021) for empirical systems, we determined the population-level group-size distribution $p_g$. This distribution is derived from $k_g$, defined in their work as the average number of hyperlinks of order $g$ a node is involved in, calculated as $k_g = gL_g/N$. The proportion $p_g$ is then $k_g$ normalized by the sum of all such average $g$-hyperdegrees, $k_{total} = \sum_x xL_x/N$. Therefore, the population-level group-size distribution, which sums to 1 (i.e., $\sum_g p_g = 1$), is given by Eq 2. This distribution $p_g$ reflects the typical collaborative engagement within each field, weighted by group size.

$$p_g = \frac{\frac{gL_g}{N}}{\frac{\sum_x xL_x}{N}} = \frac{gL_g}{\sum_x xL_x} \qquad (2)$$

Adopting the game-theoretic approach of Alvarez-Rodriguez et al. (2021), we assume the observed collaboration structure factor $r(g)$ is extracted by assuming a proportional relationship $r(g) = zp_g$. The equilibrium condition for the Public Goods Game is:

$$\sum_g p_g(1 - r(g)) = 0 \qquad (3)$$

Given $\sum_g p_g = 1$, substituting $r(g) = zp_g$. This yields the empirical reduced synergy factor $r(g)_{emp} = zp_g$ for each discipline. The empirical synergy factor $R(g)_{emp}$ is then calculated as:

$$R(g)_{emp} = r(g)_{emp} \cdot g \qquad (4)$$

In scientific collaboration, this assumption $(r(g) \propto p_g)$ posits that scientists, on average, rationally allocate their collaborative efforts towards team sizes that offer a perceived optimal balance of return (e.g., publication impact, knowledge gain, career progression) (Wu et al., 2024) against the associated costs (e.g., coordination time, resource sharing, potential for conflict) (Wu et al., 2024). Therefore, a higher population-level participation frequency $r(g)$ experienced or anticipated by the community from that particular collaborative configuration. We then define the synergy factor as $R(g)_{emp} = r(g)_{emp} \cdot g$, reflecting the total collaborative payoff achievable by a group of size $g$. This yields empirical curve $R(g)_{emp}$ for each discipline, representing the effective benefit derived from group interactions of varying sizes. To elucidate the mechanisms shaping the empirical $R(g)_{emp}$ curve, we employed the cost-benefit decomposition model proposed by Alvarez-Rodriguez et al. (2021). This model posits that the synergy factor arises from two competing processes: a benefit term associated with increasing productivity with group size, and a cost term reflecting coordination overhead. The theoretical synergy factor $R(g)_{model}$ is formally expressed as:

$$R(g; \alpha, \beta, \gamma) = \alpha \cdot g^\beta \cdot e^{-\gamma(g-1)} \qquad (5)$$

Here, $\alpha$ is a scaling constant, $\beta$ controls the strength of increasing returns to scale, and $\gamma$ quantifies the exponential growth of organizational costs. The model has an interior maximum when $\frac{\beta}{\gamma} > 1$, indicating an optimal group size that balance benefits and costs.

Following Alvarez-Rodriguez et al. (2021), the parameter $\alpha$ in Eq5 is not treated as an independent free parameter but is determined by the critical point condition (Eq3). The model`s reduced synergy factor is $r(g)_{model} = \frac{R(g;\alpha,\beta,\gamma)}{g} = \alpha \cdot g^{\beta-1} \cdot e^{-\gamma(g-1)}$. Substituting this into Eq3, and using $\sum_g p_g = 1$, $\alpha$ is defined for any given $\beta$ and $\gamma$ as: $\alpha(\beta,\gamma) = 1/\sum_g p_g g^{\beta-1} \cdot e^{-\gamma(g-1)}$. The parameters $(\beta,\gamma)$ were estimated for each discipline by performing a non-linear least-squares fitting produced. This involved minimizing the residual sum of the square (RSS) between the empirical synergy factor $R(g)_{emp}$ from Eq4 and the model`s prediction $R(g)_{model}$ (from Eq5, with $\alpha$ calculated via $\alpha(\beta,\gamma) = 1/\sum_g p_g g^{\beta-1} \cdot e^{-\gamma(g-1)}$):

$$RSS = \sum_i \left[ R(g_i)_{emp} - \alpha(\beta,\gamma) \cdot g_i^{\beta} \cdot e^{-\gamma(g_i-1)} \right]^2 \tag{6}$$

The minimization was performed with respect to $\beta$ and $\gamma$. In each iteration of this optimization process, the value of $\alpha$ was dynamically calculated using the current values of $\beta$, $\gamma$, and the empirically determined $p_g$ distribution. To ensure the robustness of the fit and mitigate the risk of local minima, the optimization was initiated from multiple starting points for $\beta$ and $\gamma$. Based on typical expectations, initial values for $\beta$ were explored in the range of 0 to 2, and for $\gamma$ in a range such as 0.01 to 0.5. During the optimization, constraints $\beta \geq 0$ and $\gamma \geq 0$ were strictly enforced. The parameter $\alpha$ is inherently positive due to its definition in $\alpha(\beta,\gamma) = 1/\sum_g p_g g^{\beta-1} \cdot e^{-\gamma(g-1)}$), assuming non-negative $p_g$, $g$, and $e^{-\gamma(g-1)}$ terms. For improved reliability of the fitting process, group sizes $g$ for which the total number of publications $L_g$ was fewer than 100 in a given discipline were excluded from the sum in Eq6. The performance of the fitted model for each discipline was evaluated using the coefficient of determination, $R^2$. The fitted parameters $(\alpha,\beta,\gamma)$ allow for the classification of disciplines based on their distinct collaborative characteristics, such as the extent of increasing returns to scale (indicated by $\beta$) and the burden of coordination costs (indicated by $\gamma$).

**Mediation and moderation analyses**

To investigate the mechanisms linking team size to scientific disruption, we performed mediation and moderation analyses. These analyses were separately for each of the 19 top-level fields.

**Mediation analysis**

We first examined whether discipline-specific synergy factor ($R(g)$), mediates the relationship between team size ($g$) and the disruption index (DI). The variables for the mediation model were defined as follows. The independent variable was team size ($g$), corresponding to the number of authors on a publication. The mediator was the synergy factor ($R(g)$), operationalized as the value derived from the fitted function in Equation 6 for the specific team size ($g$) of each publication, using the $\alpha$, $\beta$ and $\gamma$ parameters previously estimated for its respective discipline. The dependent variable was the disruption index (DI) of the publication, calculated according to Equation 1. The mediation analysis followed a path analytic framework. We estimated (1) the total effect of team size on disruption (path c: $g \rightarrow$ DI); (2) the effect of team size on the synergy factor (path a: $g \rightarrow R(g)$); and (3) the effect of the synergy factor on disruption while controlling for team size (path b: $R(g) \rightarrow$ DI | $g$) (Supplementary Note 1).

**Moderation analysis**

We investigate whether higher-order team structural characteristics moderate the relationship between synergy factor $R(g)$ and DI. For each publication, the team (hyperlink) was characterized by six features, which served as potential moderators ($W$) in our models: team academic age variance, team historical productivity variance, team historical citation variance, team historical average DI variance, team disciplinary experience variance, team gender proportion. The detailed descriptions are in Supplementary Note 2. For each of the six moderators, we conducted a separate moderation analysis by fitting a linear regression model for each top-field. The model predicted the DI using $g$ as a control, $R(g)$, $W$, and an interaction term between the synergy factor and moderator ($R(g) \times W$):

$$DI = \beta_0 + \beta_1 g + \beta_2 R(g) + \beta_3 W + \beta_4 (R(g) \times W) + \varepsilon \qquad (8)$$

The coefficient $\beta_4$ of the interaction term ($R(g) \times W$) quantifies the moderating effect. A statistically significant $\beta_4$ indicates that the relationship between the synergy factor $R(g)$ and disruption DI varies depending on the level of the team characteristic W. We report the estimated interaction coefficients ($\beta_4$), their standard errors, and 95% CIs for each moderator across the 19 disciplines. These moderating effect sizes will be visualized using bar charts, displaying the 95% CIs, with statistically significant coefficients (p < 0.05) highlighted by solid circular markers.

**Results**

**Synergy dynamics reveal discipline-specific collaboration patterns**

Since 1960, the average team size of academic papers has shown an essentially linear increase across scientific disciplines (Fig. 1b). However, examining specific disciplines revealed remarkable heterogeneity in growth patterns and magnitudes. Physics exhibited the most dramatic transformation, with team sizes growing from 1.66 authors in 1960 to 7.95 authors in 2020—a 350.56% increase that reflected the rise of large-scale experimental collaborations in high-energy physics and cosmology. Biology and Medicine followed with substantial growth rates of 223.32% and 214.77%, respectively, consistent with the increasing complexity of life sciences research and multi-institutional clinical trials. In contrast, humanities disciplines maintained minimal growth or even decline (Fig. S1). Art showed a slight decrease (-0.05%), while Philosophy and History demonstrated modest increases of only 13.06% and 15.19%, respectively, reflecting the traditionally individualistic nature of scholarship in these fields (Fig. 1c). Several STEM disciplines exhibited intermediate patterns: Chemistry and Materials Science showed steady growth (Chemistry: 2.03 to 5.46 authors; Materials Science: 1.75 to 4.85 authors), while Computer Science demonstrated rapid acceleration after 1990, coinciding with the field`s expansion and increasing interdisciplinary connections (Fig. S1).

Based on these observed collaboration patterns, we inferred discipline-specific and time-specific reduced synergy factors $R(g)$, which reflect the effective collaborative payoff for a group of size $g$ within a given context. The shapes of these empirical $R(g)$ curves, representing the net benefits of collaboration as a function of team size, varied considerably across disciplines (Fig. 1d, Fig. S2) and over time (Fig. 1e). Physics showed a characteristic pattern with peak synergy for small to medium team sizes ($R(3) \approx 8.5$, $R(4) \approx 9.1$) before gradually declining for larger groups, suggesting that while Physics has embraced increasingly large collaborations, marginal benefits per additional member diminish significantly beyond small core teams. Conversely, Philosophy exhibited its highest $R(g)$ value at $g = 1$ ($R(1) \approx 1.32$) with sharp decreases for $g > 1$, indicating maximal synergy for solo work and rapidly diminishing returns for collaboration. Other disciplines displayed distinct intermediate patterns: Chemistry ($R(5) \approx 6.4$) and Biology ($R(6) \approx 7.7$) showed optimal synergy at larger team sizes compared to Philosophy but with different peak points and decay rates compared to Physics. Medicine exhibited a broader optimal range with peak synergy at group sizes of 5-6 members ($R(6) \approx 8.6$), reflecting the collaborative nature of clinical research requiring diverse expertise. Temporally, the overall $R(g)$ curves (aggregated across disciplines) also evolved significantly. The $R(g)$ values for larger team sizes became relatively more pronounced in later decades, with $R(g)$ for $g = 30$ increasing from approximately 1.49 in 1960 to 6.52 in 2020, indicating enhanced viability of large-scale collaborations over time (Fig. 1e).

To further dissect these collaborative dynamics, we fitted the function 5 to quantify the "economics of collaboration." The estimated parameters revealed distinct collaboration archetypes across disciplines. **Experimental sciences** such as Geology ($\beta \approx 2.54$, $\gamma \approx 0.58$) and Materials Science ($\beta \approx 3.69$, $\gamma \approx 0.70$) exhibited high $\beta$ values, suggesting strong increasing returns to scale, coupled with moderate coordination costs. **Humanities** including Philosophy ($\beta \approx 0$, $\gamma \approx 0.77$) and Art ($\beta \approx 0$, $\gamma \approx 0.47$) showed negligible scaling benefits from collaboration, with synergy primarily determined by α and diminishing with coordination costs. **Physics** presented a unique profile with $\beta \approx 0$ but very low $\gamma \approx 0.03$, indicating that while initial synergy doesn`t scale strongly with group size, coordination costs grow very slowly, allowing larger teams to remain viable. (Fig. 1f, g). During 1960-2020, the synergy parameters themselves evolved. The benefit parameter $\beta$ showed a gradual decline from approximately 2.09 in 1960 to 2.06 in 2020, suggesting that while team sizes have grown, marginal benefits of additional collaboration have decreased. Conversely, the cost parameter $\gamma$ decreased over the same period, indicating that coordination penalties have become less severe, likely due to improved communication technologies and institutional support for large-scale collaboration (Fig. 1g).

Principal component analysis of the synergy parameters ($\alpha, \beta, \gamma$) reveals three distinct clusters of disciplinary collaboration patterns that remain stable across decades. **Cluster 1** comprised experimental sciences (Physics, Chemistry, Materials Science, Medicine, Biology) characterized by substantial synergy benefits for medium-sized groups and moderate coordination costs. **Cluster 2** included applied and computational fields (Engineering, Computer Science, Environmental Science) with intermediate synergy profiles and balanced benefit-cost ratios. **Cluster 3** encompassed humanities and social sciences showing minimal synergy benefits beyond individual collaborations and high coordination costs for larger groups.

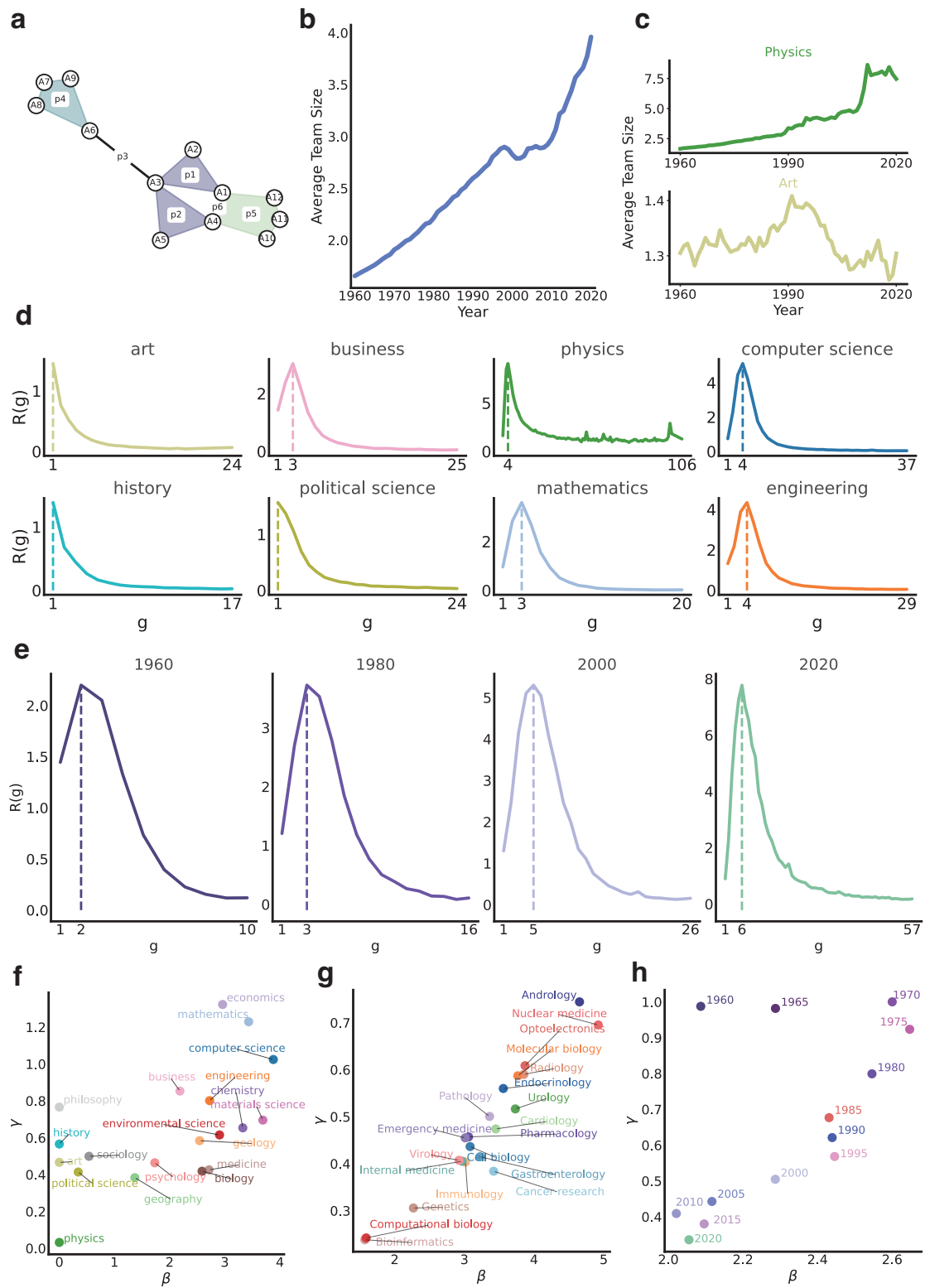

**Fig. 1. Team size and $R(g)$ distribution across disciplines and time. a,** Hypergraph schematic diagram. $p_1, p_2, p_3, \ldots, p_n$ refer to hyperedges. $A_1, A_2, \ldots, A_n$ refer to the nodes of the hyperedges (here referring to the authors of papers). **b,** Overall average team size from 1960-2020. **c,** Comparison between 2020 and 1960, showing the discipline with the largest increase in average team size (Physics) and the smallest (Art). **d,** $R(g)$ values corresponding

to each team size across eight disciplines, with dashed lines indicating the team size corresponding to maximum synergy ($R(g)$ peak) for each discipline. All fitted curves have R² values greater than 0.9. **e,** Temporal development trends of $R(g)$. Dashed lines indicate the team size corresponding to maximum synergy ($R(g)$ peak) for each year. All fitted curves have R² values greater than 0.9. **f,** When fitting each $R(g)$ curve, we obtained the $\beta$ and $\gamma$ values for each discipline. $\beta$ controls the strength of the benefits that come with increasing team size, known as "increasing returns to scale." A higher $\beta$ suggests that the group becomes more productive or efficient as it grows. $\gamma$ quantifies how quickly organizational costs grow as the team gets larger. A higher $\gamma$ implies that managing a larger group becomes more difficult and costly at a faster rate. The model has an optimal group size only when the condition $\beta/\gamma > 1$ is satisfied. Based on the fitting results, disciplines with $\beta/\gamma > 1$ include Physics, Arts, History, Philosophy, Sociology, and Political Science. **g,** Top 20 subfields ranked by $\beta/\gamma$ values among all subfields. **h,** Temporal distribution of $\beta$ and $\gamma$ values by year.

## Synergy mediates team size effects with Discipline-specific heterogeneity moderation

To understand the mechanisms through which team size influences research disruption, we conducted mediation analysis examining whether the synergy factor $R(g)$ served as an intermediary pathway. Our analysis across 19 scientific disciplines revealed that the synergy factor functions as a crucial mediator, with significant mediation effects observed in 18 of 19 disciplines (94.7%, Table S1). The mediation analysis demonstrated substantial heterogeneity across disciplines in how team size influences disruption. Total effects of team size on disruption were significant in 89.5% disciplines, with effect sizes ranging from minimal in Philosophy (estimate = -0.003, $P$ = 0.394) to substantial in Geography (estimate = -0.102, $P$ < 0.001) and Environmental Science (estimate = -0.099, $P$ < 0.001). Notably, direct effects—representing the impact of team size on disruption when controlling for the synergy factor—were significant across 18 disciplines (94.7%), indicating that team size influences disruption through both direct pathways and indirect pathways mediated by collaborative synergy (Table S1). The indirect effects, representing the pathway through which team size influences disruption via the synergy factor, were significant in 94.7% disciplines. Computer Science exhibited the strongest mediation (74.8% of total effect), followed by Physics (81.5%) and Biology (84.4%), suggesting that in these fields, team collaborative synergy is particularly critical for determining research outcomes (Fig. 2). Intriguingly, several disciplines—including Art, History, Political Science, and Sociology—showed suppression effects, where the indirect effect operated in the opposite direction to the direct effect.

To examine how team composition characteristics influence the effectiveness of collaborative synergy, we analyzed six team heterogeneity dimensions as moderators of the relationship between synergy factor $R(g)$ and disruption. Age heterogeneity emerged as the most consistent moderator, showing significant moderation effects in 89.5% disciplines. The direction of moderation varied systematically across disciplinary contexts: life sciences fields (Biology, Medicine, Psychology) consistently showed positive moderation effects (estimate = 0.026, 0.009, and 0.033, respectively; all $P < 0.001$), suggesting that age diversity enhanced the benefits of collaborative synergy in these domains. Conversely, Political Science and Sociology exhibited negative moderation effects (estimate = -0.011 and -0.005, respectively; $P < 0.001$), indicating that age diversity may impede effective collaboration in social science contexts (Fig. 2). Productivity heterogeneity demonstrated significant moderation in 77.8% disciplines, with generally positive effects in empirical sciences. Geography showed the strongest positive moderation (coef = 0.016, $P < 0.001$), followed by Business (coef = 0.009, $P < 0.001$) and Geology (coef = 0.011, $P < 0.001$). These findings suggested that complementary productivity levels within teams enhance collaborative effectiveness, particularly in fields requiring diverse methodological approaches (Fig. S3).

Interdisciplinary heterogeneity exhibited significant moderation effects in 84.2% disciplines (), with the strongest positive effects observed in Geography (coef = 0.040, $P < 0.001$), Business (coef = 0.031, $P < 0.001$), and Psychology (coef = 0.035, $P < 0.001$). This pattern indicates that interdisciplinary collaboration enhances synergy effects, consistent with the growing importance of cross-disciplinary research in addressing complex scientific problems (Fig. S3). DI heterogeneity showed significant moderation in 66.7% disciplines. Notably, the direction of effects varied substantially: positive moderation in Geology (coef = 0.022, $P < 0.001$) and Geography (coef = 0.012, $P < 0.001$), but negative moderation in Biology (coef = -0.004, $P < 0.001$) and Political Science (coef = -0.005, $P = 0.002$). This suggested that while cognitive diversity can enhance collaborative synergy in some contexts, it may introduce coordination challenges in others (Fig. 2). Citation heterogeneity presented more mixed results, with significant moderation in only 50.0% discipline. Materials Science showed the strongest positive moderation (coef = 0.012, $P < 0.001$), while Chemistry exhibited negative moderation (coef = -0.007, $P < 0.001$), indicating that the benefits of reputation diversity are highly field-dependent (Fig. S3). Gender proportion showed significant moderation effects in 68.4% disciplines, with modest effect sizes across fields. The direction varied by discipline, with positive effects in Political Science (coef = 0.016, $P < 0.001$) and Sociology (coef = 0.013, $P < 0.001$), but negative effects in Psychology (coef = -0.004, $P < 0.001$) and Biology (coef = -

0.007, *P* < 0.001) (Fig. 2).

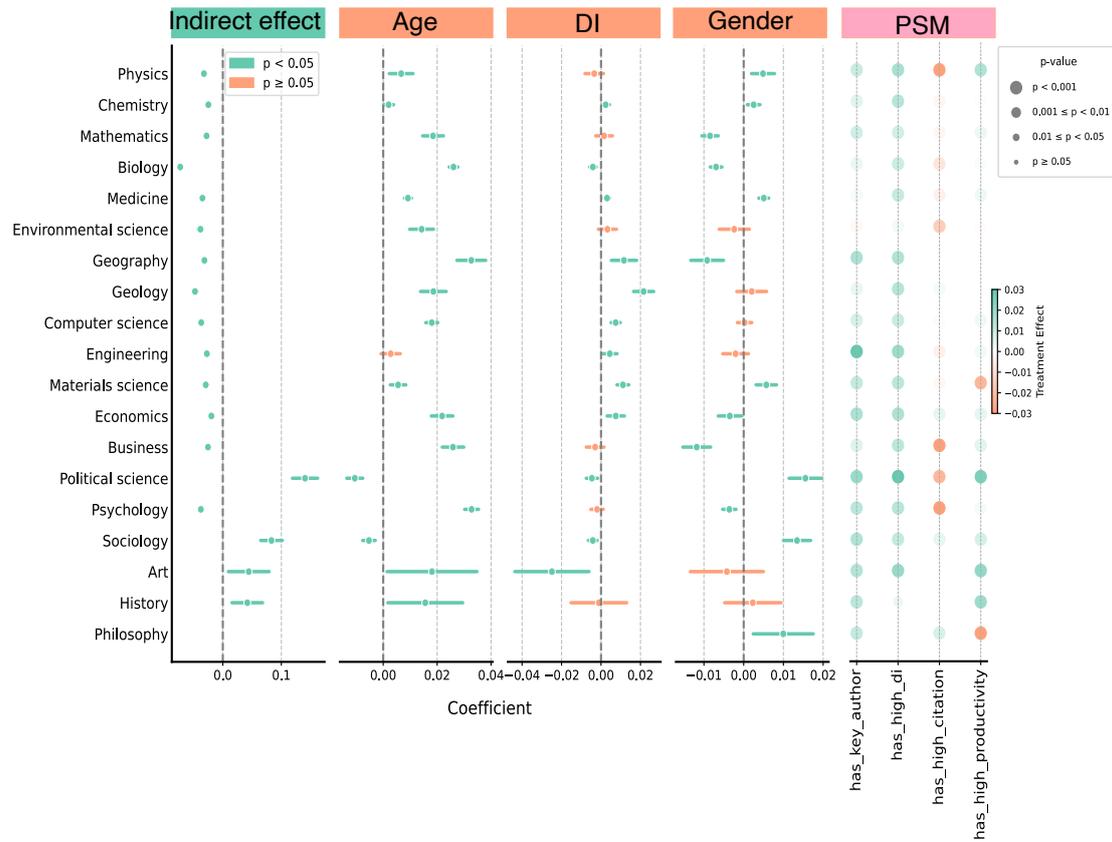

**Fig 2. Mediation effects of $R(g)$ and moderation effects of team structural characteristics.** The first column tests $R(g)$ as a mediator variable to examine whether it mediates the relationship between team size and DI through path analysis. Results show that four disciplines—Art, History, Political Science, and Sociology—showed suppression effects, where the indirect effect operated in the opposite direction to the direct effect. The second to fourth columns examine the moderating effects of team age heterogeneity, DI experience heterogeneity (i.e., the heterogeneity of average DI values across all papers published by team members prior to the focal paper), and team gender composition (values closer to 1 indicate more male members in the team) on the relationship between team size and DI. The last column shows the treatment effects of key authors on DI. Key authors are classified into three types: high DI experience authors, highly cited authors, and high productivity authors. An author is marked as a key author when their DI experience, cumulative citations, or publication count reaches the top 10% within their respective discipline. When conducting PSM analysis, we performed separate analyses within each discipline to examine the treatment effects of having key authors (independent variable: has_key_author), having high DI experience authors (independent variable: has_high_di), having highly cited authors (independent variable: has_high_citation), and having high productivity authors (independent variable:

has_high_productivity) on paper DI (dependent variable). All experiments controlled for publication time, team size, $R(g)$, age heterogeneity, DI experience heterogeneity, gender composition, interdisciplinary degree heterogeneity, and productivity heterogeneity. Orange bubbles in the figure represent negative effects, green bubbles represent positive effects, lighter bubbles indicate effects closer to zero, and larger bubbles indicate more significant p-values.

**Key authors drive DI through position-dependent mechanisms**

While previous studies have examined the relationship between team composition and scientific impact, the causal role of key authors—those with exceptional citation records, productivity, or disruptive capacity—in fostering breakthrough research remains unclear. To address this question, we employed propensity score matching (PSM) to identify the causal effects of key authors on scientific disruption, controlling for team diversity in age, productivity, citation patterns, disruptive capacity, gender composition, and team size. Our analysis revealed that key authors play a fundamental role in driving scientific disruption across academic disciplines. Among the 29.3% of papers with key authors in our dataset, we observed substantial heterogeneity in author types: 16.1% feature high-citation authors, 15.5% include high-productivity authors, and 16.6% contain high-disruption authors. This distribution suggested that different types of expertise contribute to research teams in distinct ways (Fig. 2).

We found that the presence of key authors significantly enhances scientific disruption (treatment effect = 0.0149, $P < 0.001$) after controlling confounding factors. Papers with key authors have 561.4% higher DI than thoses without key authors (Fig 3a), supporting the hypothesis that exceptional researchers catalyze breakthrough discoveries. However, our analysis reveals nuanced differences across author types that challenge conventional assumptions about scientific excellence (Table. S2). Surprisingly, while the presence of high-citation authors—traditionally viewed as markers of scientific quality—correlated with reduced disruption (treatment effect = -0.0150, $P < 0.001$), high-disruption authors demonstrate the strongest positive effect on team innovation (treatment effect = 0.0152, $P < 0.001$). High-productivity authors show a modest but significant positive contribution (treatment effect = 0.0009, $P < 0.001$). These contrasting effects suggested that citation-based metrics may not capture the research qualities most conducive to breakthrough science, aligning with recent critiques of traditional academic evaluation systems (Table. S2).

The organizational structure of research teams significantly influenced how key authors contribute to scientific disruption. Our analysis of author position revealed that key authors in first-author positions generate substantially higher disruption (mean DI = 0.0204) compared to those in non-first positions (mean DI = 0.0092) (Fig 3b). This 121% difference suggested that placing key authors in leadership roles maximizes their innovative potential. Conversely, key authors in last author positions showed similar disruption levels to non-last authors (mean DI = 0.0093 vs. 0.0136), indicating that supervisory roles may not effectively leverage exceptional talent for breakthrough research (Fig 3c). Key authors in middle positions demonstrated intermediate performance (mean DI = 0.0110), suggesting that central coordination roles provide moderate advantages for disruptive innovation (Fig. 3d).

The impact of key authors varies dramatically across academic disciplines, revealing important contextual dependencies in how exceptional talent influences scientific progress. Engineering exhibited the strongest overall effect (treatment effect = 0.0385), followed by Political Science (0.0311 for high-disruption authors) and Art (0.0211 for high-productivity authors). These field-specific variations suggested that the organizational and methodological characteristics of different disciplines create varying opportunities for key authors to drive innovation (Table S2). Notably, STEM fields showed more modest effects despite their collaborative nature and resource intensity. This pattern may reflect differences in research paradigms, with more established fields exhibiting greater resistance to disruptive approaches, while emerging or interdisciplinary areas provide more fertile ground for breakthrough discoveries (Table S2).

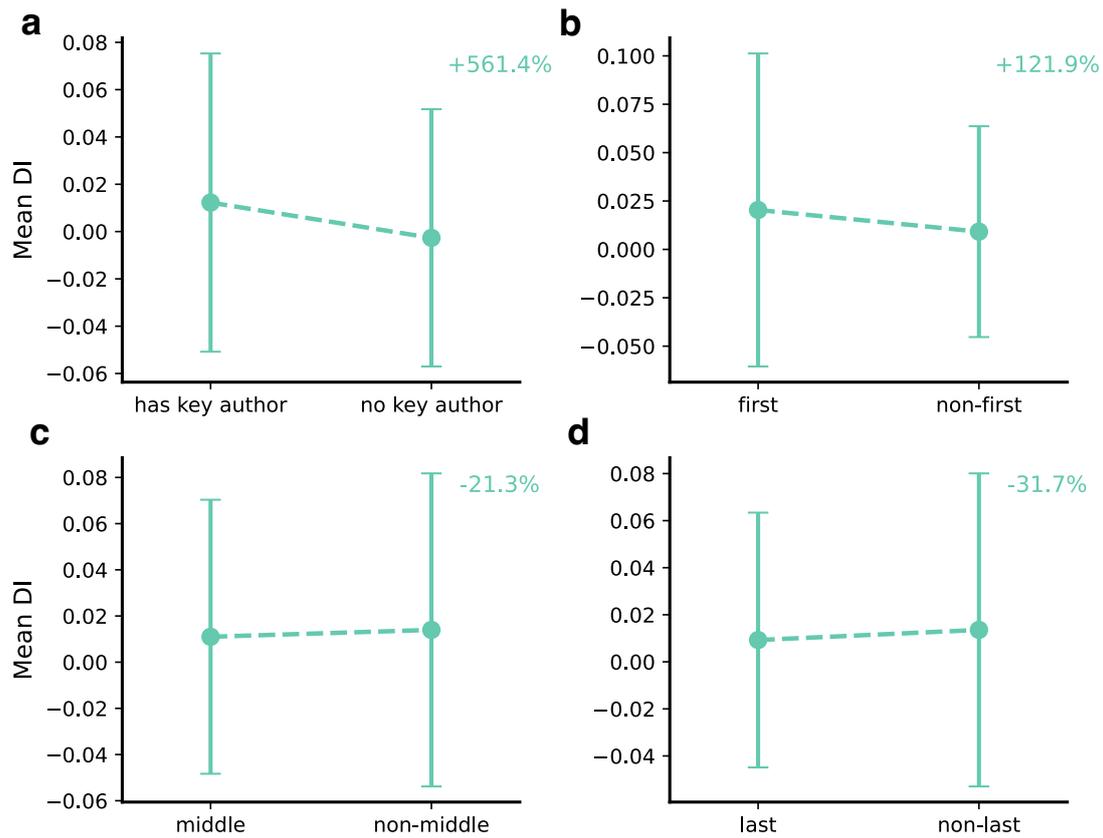

**Fig 3. Impact of key author`s position on DI. a,** Papers with key authors have 561.4% higher DI than papers without key authors. **b,** Among papers with key authors, papers where key authors serve as first authors have 121.9% higher DI than papers where key authors serve as non-first authors. **c,** Among papers with key authors, papers where key authors serve as middle authors have 21.3% lower DI than papers where key authors serve as non-middle authors. **d,** Among papers with key authors, papers where key authors serve as last authors have 31.7% lower DI than papers where key authors serve as non-last authors.

**Four knowledge production modes shape scientific collaboration and innovation**

To systematically understand the intrinsic mechanisms of scientific collaboration, we integrated the core dimensions from our previous analyses—team size, synergy factor parameters, key author characteristics, and team heterogeneity indicators—to construct a multi-dimensional feature space comprising 22 variables. Through a clustering optimization strategy combining the elbow method and silhouette coefficient analysis, we determined the optimal number of clusters to be four. Our analysis identified four distinctly different knowledge production modes, each embodying unique collaborative mechanisms and theoretical characteristics (Figure 4a). The **elite-driven mode** comprised 16.5% of the sample and was characterized by extremely high key author density (83.5%) and the highest citation impact (average of 38.9 citations), validating the significant role of key author effects in small teams (1.66 members). The **baseline mode** constituted the largest proportion (44.5%) and represented the standardized scientific collaboration pattern, with all indicators at moderate levels, providing an important comparative baseline for other modes. The **heterogeneity-driven mode** accounted for 35.3% and exhibited the highest level of team heterogeneity and synergy efficiency ($R(g) = 5.134$), embodying the mechanism by which diversified large teams (4.64 members) drive innovation through heterogeneity. The **low-cost mode**, while representing the smallest proportion (3.7%), featured extremely low coordination costs ($\gamma = 0.032$), representing the ideal state of efficient elite coordination.

Different knowledge production modes demonstrated significant differences across three core output dimensions (all Kruskal-Wallis tests p < 0.001). Regarding disruption indices, the elite-driven mode performed most prominently, which corresponded with its extremely high key author concentration, supporting the hypothesis of high-impact researchers` crucial role in breakthrough discoveries (Fig. 4b). Citation analysis revealed a more complex pattern: the elite-driven mode achieved the highest average citations (38.9), followed by the heterogeneity-driven mode (31.5), while the baseline and low-cost modes received 24.5 and 28.0 citations respectively. This finding indicated that different collaborative mechanisms orient toward different types of scientific impact—elite-driven modes optimize impact maximization, while heterogeneity-driven modes achieved balanced high-impact output through diversification (Fig. 4c). Analysis of atypical combination (Uzzi et al., 2013) further revealed the innovation characteristic differences among modes (Fig. 4d). The heterogeneity-driven mode excelled in the novelty dimension, closely related to its highest cross-disciplinary heterogeneity, validating the advantages of diverse teams in knowledge recombination and conceptual innovation. In contrast, the baseline mode performed relatively conservatively in novelty, reflecting the characteristic that standardized collaboration modes tend toward incremental development rather than breakthrough innovation.

The preferences of different disciplines for various knowledge production modes exhibited strong field-specific characteristics, reflecting the profound influence of different research paradigms and knowledge production traditions (Fig. 4e). Physics showed extreme mode concentration, with 99.98% of papers adopting the low-cost mode, highly consistent with physics` big science project traditions and highly standardized experimental collaboration patterns. In contrast, humanities disciplines such as Art (86.25%), History (81.87%), and Philosophy (81.45%) clearly favored the baseline mode, reflecting the traditional characteristics of individual scholar dominance and standardized collaboration in humanities research. Life sciences fields demonstrated strong preference for the heterogeneity-driven mode: Biology (49.64%), Medicine (48.10%), and Environmental Science (49.28%) all had nearly half their papers adopting this mode, reflecting the increasingly complex and interdisciplinary trends in life sciences research and the need for integrating diverse professional knowledge (Fig. 4e). Materials Science (49.33%) and Chemistry (50.92%) similarly showed preference for the heterogeneity-driven mode, reflecting these fields` emphasis on cross-disciplinary knowledge fusion in technological innovation (Fig. 4e). Engineering and Computer Science displayed relatively balanced mode distributions, possibly reflecting these application-oriented disciplines` need to select different collaboration strategies based on specific problem characteristics. Notably, the elite-driven mode maintained a relatively stable proportion (12-21%) across all disciplines, suggesting the existence of key author effects as a universal scientific phenomenon (Fig. 4e).

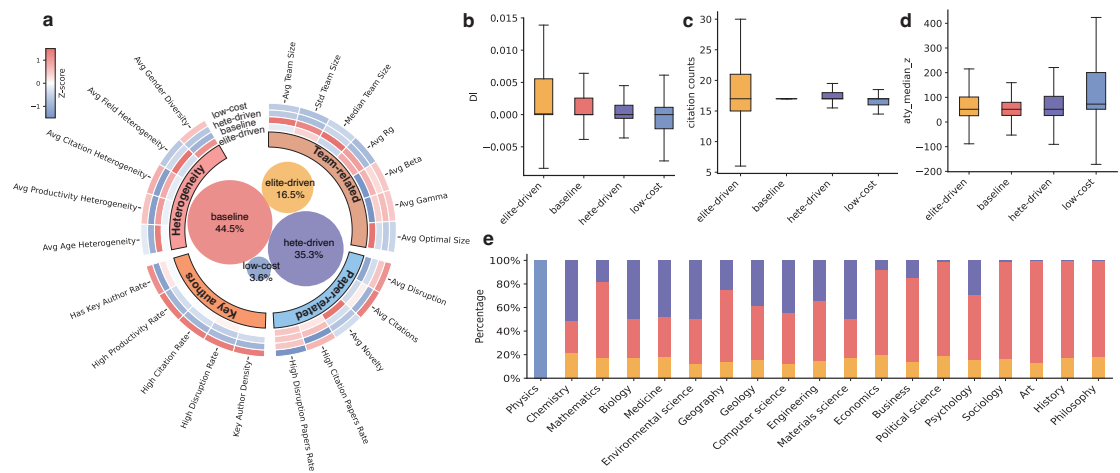

**Fig. 4. The classification and distribution of four knowledge production modes. a,** Circular heatmap of knowledge production mode profiles. It presents a circular heatmap visualizing the results of our clustering analysis on research teams using 22 features across four core dimensions: heterogeneity, paper-related indicators, team-related indicators, and key authors. Through this comprehensive analysis, we identified 4 distinct knowledge production modes in scientific collaboration. The baseline mode represents the largest proportion at 44.5% of the sample, followed by the heterogeneity-driven mode (hete-driven mode) at 35.3%, the elite-driven mode at 16.5%, and the low-cost mode with the smallest share at 3.6%. To enable in-depth analysis and intuitive comparison of each mode's characteristics, we applied Z-score standardization to all features, eliminating scale differences between variables and ensuring fair cross-feature comparisons. In the visualization, the Z-score values have clear interpretations: positive values (shown in red) indicate that a cluster mode performs above the overall average on specific features, representing its "strengths"; negative values (shown in blue) indicate

below-average performance, representing its "weaknesses"; and values close to zero (shown in white) represent performance comparable to the overall average. **b,** The DI distribution of the four knowledge production modes. **c,** The citation distribution of the four knowledge production modes. **d,** The aty_median_z distribution of the four knowledge production modes. This value represents the degree of conventionality of the target paper in its reference combination. When this value is less than 0, it means the paper is atypical, while values greater than 0 indicate the paper is conventional. **e,** The disciplinary distribution of the four knowledge production modes.

Over the past six decades, the distribution of knowledge production modes has undergone significant historical transitions, profoundly reflecting the evolution of scientific research organization (Fig. S4). In the 1960s, the elite-driven mode dominated (59.0%), reflecting the central role of individual scholars or small elite teams in traditional scientific research. However, this proportion continuously declined over subsequent decades, reaching only 10.4% in the 2020s, marking a fundamental transition from the "heroic scientist" era to the collaborative science era. The trajectory of the baseline mode was equally noteworthy: rising from 30.52% in the 1960s to peak at approximately 50% during the 1970s-2000s, then gradually declining to 33.73% in the 2020s. This inverted-U trend may reflect scientific research`s evolution through standardization and institutionalization phases before progressing toward more complex and diversified collaboration modes. The most significant change was the sustained growth of the heterogeneity-driven mode: from merely 5.43% in the 1960s to a dramatic rise to 52.20% in the 2020s, becoming the dominant mode of contemporary scientific research. This trend highly aligned with the increasing complexity of scientific problems, enhanced interdisciplinarity, and the advent of the big science era. The accelerated growth after 2000 (from 34.81% to 52.20%) may be closely related to digital technologies reducing collaboration costs, globalization promoting international cooperation, and funding agencies' policy support for interdisciplinary research. The low-cost mode maintained a relatively stable small proportion (3-5%) throughout the period, reflecting the scarcity and special value of this efficient coordination mechanism. Its relatively high proportion in the 1960s (5.05%) may reflect the characteristic that smaller early scientific communities more easily achieved low-friction coordination (Fig. S4).

**Discussion**

**The Economics of Scientific Collaboration**

The discipline-specific synergy curves $R(g)$ provided unprecedented insight into the "sweet spots" of collaboration across fields. Physics's peak synergy at medium team sizes ($R(3) \approx 8.5$, $R(4) \approx 9.1$) reflectd the field's evolution toward large-scale experimental collaborations while maintaining the cognitive benefits of smaller core teams (Chu et al., 2025). Conversely, Philosophy's maximal synergy at g=1 ($R(1) \approx 1.32$) with sharp decay for larger groups validated the traditional emphasis on individual scholarship in humanities (An & Shan, 2023). These patterns suggest that optimal collaboration structures are not universal but emerge from the interplay between disciplinary knowledge characteristics, methodological requirements, and institutional contexts (Becher & Trowler, 2001; Hilton & Cooke, 2015).

The $\beta$ and $\gamma$ parameters from our synergy factor analysis illuminated the fundamental economic principles that govern the collective behavior of scientific teamwork. The $\beta$ parameter, representing scaling benefits, captured how additional team members contribute to research productivity, while $\gamma$ reflectd coordination costs that increase with team size (Alvarez-Rodriguez et al., 2021). The temporal evolution showing decreased marginal benefits ($\beta$: 2.09→2.06) coupled with reduced coordination costs ($\gamma$ decline) suggested that technological advances in communication and collaboration tools have fundamentally altered the economics of scientific teamwork (Brucks & Levav, 2022; Cummings & Kiesler, 2005).

The emergence of three stable collaboration archetypes—experimental sciences with high $\beta$ and moderate $\gamma$, applied fields with intermediate profiles, and humanities with minimal $\beta$ and high $\gamma$—represented distinct evolutionary solutions to the collaboration optimization problem. These archetypes aligned with Whitley's (2000) taxonomy of scientific fields but provided quantitative characterization of their collaborative dynamics. The stability of these clusters across decades indicated that disciplinary collaboration patterns represent deep structural features rather than transient adaptations, consistent with theories of institutional persistence in scientific communities (Kuhn, 1962).

**Synergy as the Missing Link in Team Size-Disruption Relationships**

Our mediation analysis resolved a fundamental paradox: how larger teams can simultaneously be associated with reduced disruption while enabling complex, high-impact discoveries (Wu et al., 2019). The finding that synergy factor $R(g)$ mediated 75-85% of team size effects on disruption across most disciplines demonstrated that collaborative effectiveness, not mere scale, determined innovative outcomes. This mechanistic insight explained why previous studies found inconsistent team size-innovation relationships without accounting for the quality and efficiency of collaborative processes (Giffoni et al., 2025; Wang & Barabási,

2021; Wuchty et al., 2007).

The suppression effects observed in Art, History, Political Science, and Sociology—where indirect effects operate opposite to direct effects—revealed that collaboration may actually inhibit breakthrough thinking in fields requiring deep individual reflection and conceptual innovation. These findings aligned with research on the tension between collaboration and creativity, particularly in knowledge domains where synthesis and interpretation were paramount (Csikszentmihalyi, 1997; Simonton, 2004). The discipline-specific nature of these effects underscored the importance of context-dependent models of scientific innovation.

**Team Composition Heterogeneity as Synergy Modulators**

Age heterogeneity emerged as the most consistent moderator of synergy effects (89.5% of disciplines), but with striking directional differences across fields. The positive moderation in life sciences (Biology: 0.026, Medicine: 0.009, Psychology: 0.033) versus negative effects in social sciences (Political Science: -0.011, Sociology: -0.005) suggested that optimal diversity levels depend on disciplinary task characteristics and collaboration requirements. This pattern aligned with diversity research showing that the benefits of heterogeneity depend on task complexity and the nature of knowledge integration required (Modi et al., 2025; van Knippenberg, 2024).

The strong positive moderation effects of interdisciplinary heterogeneity (84.2% of disciplines) validated the increasing importance of cross-disciplinary research for addressing complex scientific problems (Yang, 2025; Yu et al., 2024). Fields like Geography (0.040), Business (0.031), and Psychology (0.035) showed particularly strong benefits from interdisciplinary collaboration, reflecting their inherently integrative nature and the need to synthesize knowledge across domain boundaries. These findings supported policy initiatives promoting interdisciplinary research while highlighting the need for institutional support to manage the coordination challenges of cross-disciplinary collaboration (Rhoten & Parker, 2004).

**Key Authors as Catalysts of Scientific Disruption**

The 561% higher DI for papers with key authors provided compelling evidence for the fundamental role of exceptional talent in breakthrough science, extending beyond previous work on star scientists (Betancourt et al., 2023; Yadav et al., 2023). However, our finding that high-citation authors actually reduced disruption (-0.0150) while high-disruption authors enhancde it (0.0152) challenged traditional metrics of scientific excellence and suggested that citation-based evaluation systems may systematically undervalue disruptive contributions (Funk & Owen-Smith, 2017).

The position-dependent effects revealed crucial insights about the psychology of scientific leadership and the mechanisms of social influence. The 121% higher disruption for key authors in first-author positions demonstrated that placing exceptional talent in leadership roles maximizes innovative potential, while the lack of corresponding author benefits suggested that supervisory roles may not effectively leverage disruptive capabilities. This pattern aligned with research on scientific leadership showing that different organizational roles require different skill sets and may not equally benefit from exceptional individual talent (Xu et al., 2022).

**Knowledge Production Modes as Organizing Principles**

The four-mode typology—elite-driven (16.5%), baseline (44.5%), heterogeneity-driven (35.3%), and low-cost (3.7%)—provided a comprehensive framework for understanding the diversity of scientific collaboration strategies. Each mode represented a distinct solution to the optimization problem of balancing innovation potential, resource efficiency, and coordination costs. The elite-driven mode`s exceptional performance on disruption metrics validated the importance of star scientist effects, while the heterogeneity-driven mode's strength in novelty generation demonstrates the value of diverse expertise integration (Ottino, 2025; Uzzi et al., 2013).

The strong discipline-mode associations revealed how fields have evolved distinct collaborative cultures optimized for their specific research objectives. Physics`s extreme concentration in the low-cost mode (99.98%) reflected the field's mastery of large-scale coordination through standardized protocols and shared infrastructure (Galison, 1997). Life sciences` preference for heterogeneity-driven modes (Biology: 49.64%, Medicine: 48.10%) aligned with the increasing complexity and interdisciplinary nature of biological research (Leahey et al., 2016). These patterns suggested that successful scientific communities develop collaboration strategies that match their epistemic cultures and research challenges.

**Theoretical Implications and Paradigm Shifts**

Our findings fundamentally reconceptualized scientific collaboration from a simple scaling phenomenon to a complex optimization problem involving multiple interacting mechanisms. The synergy factor approach provided a missing theoretical layer between individual talent and scientific outcomes, offering quantitative tools for understanding how collaborative processes mediate the relationship between team composition and innovation (Fortunato et al., 2018) . This mechanistic perspective resolved apparent contradictions in previous research and provides a foundation for predictive models of scientific innovation.

The identification of distinct knowledge production modes challenged the assumption that scientific progress follows uniform patterns and instead suggests that different collaborative strategies may be optimal for different types of discoveries. This insight has profound implications for understanding the Matthew effect and cumulative advantage in science, showing how team composition can amplify or diminish individual contributions (Merton, 1968). The framework also provided new perspectives on the big science versus small science debate, demonstrating that both approaches can be optimal depending on disciplinary context and research objectives (Baumgartner et al., 2023; Price, 1963), ultimately contributing to a richer understating of how to harness collective intelligence while preserving individual creativity.

**Conclusion**

This study fundamentally reconceptualized scientific collaboration from a simple scaling phenomenon to a complex optimization problem involving synergistic dynamics, team architecture, and individual excellence. By revealing that collaborative synergy—not team size alone—mediated the relationship between collaboration and innovation, we resolved the apparent paradox of why larger teams often produce less disruptive science.

Our identification of four distinct knowledge production modes demonstrated that there was no universal optimal approach to scientific collaboration. Instead, different collaborative strategies—elite-driven, heterogeneity-driven, baseline, and low-cost modes—served different research objectives and disciplinary contexts. The finding that exceptional researchers achieve maximum impact in leadership rather than supervisory positions challenged conventional assumptions about scientific organization and highlights the importance of strategic team structure.

Several limitations must be acknowledged in interpreting these findings. The synergy factor $R(g)$ was inferred from observed collaboration patterns rather than directly measured, potentially missing nuanced collaborative dynamics such as actual knowledge exchange processes, informal collaboration networks, or the quality of intellectual contributions from different team members. The PSM approach for key author analysis, while controlling for observable confounders, cannot eliminate all selection effects. Truly comparable papers across different collaborative contexts may not exist, and unobserved factors such as research quality, institutional resources, or network effects may influence both key author participation and innovation outcomes. Additionally, our analysis focused on publication patterns rather than actual collaborative processes, limiting insights into the mechanisms of effective teamwork and knowledge integration.

**Acknowledgments**

*This research was supported by the National Social Science Fund of China "Research on the Measurement of Knowledge Contribution in Basic Research in China" (Project No. 24BTQ038) and National Experiment Base for Intelligent Evaluation and Governance, Fudan University (CXPJ2024002)*

**Declarations**

**Conflicts of interest/Competing interests**

The authors report no declarations of interest.

**Availability of data and material**

All data and materials will be made available on request.

**Code availability**

All software application or custom code support our published claims and comply with field standards.

**Ethics approval**

All of the authors appropriate approvals.

**Consent to participate (include appropriate statements)**

All of the authors consent to participate.

**Consent for publication (include appropriate statements)**

All of the authors consent for publication